\documentclass[preprint,byrevtex,balancelastpage,aps]{revtex4}%
\usepackage{amsfonts}
\usepackage{amsmath}
\usepackage{amssymb}
\usepackage{graphicx}
\usepackage{doublespace}%
\begin{document}
\title{Experimental corroboration of the Mulheran-Blackman explanation of the scale
invariance in thin film growth: the case of InAs quantum dots on GaAs(001).}
\author{M. Fanfoni,\thanks{Corresponding author, email: massimo.fanfoni@roma2.infn.it}
E. Placidi, F. Arciprete, E. Orsini, F. Patella, and A. Balzarotti}
\affiliation{Dipartimento di Fisica, Universit\`{a} di Roma "Tor Vergata", Via della
Ricerca Scientifica, I-00133 Roma, Italy.}

\begin{abstract}
Mulheran and Blackman have provided a simple and clear explanation of the
scale invariance of the island size distribution at the early stage of film
growth [Phil. Mag. Lett. \textbf{72}, 55 (1995)]. Their theory is centered on
the concept of capture zone properly identified by Voronoi cell. Here we
substantiate experimentally their theory by studying the scale invariance of
InAs quantum dots (QDs) forming on GaAs(001) substrate. In particular, we show
that the volume distributions of QDs well overlap the corrisponding
experimental distributions of the Voronoi-cell areas. The interplay between
the experimental data and the numerical simulations allowed us to determine
the spatial correlation length among QDs.

\end{abstract}
\date{}
\email{massimo.fanfoni@roma2.infn.it}
\maketitle
\preprint{ }
\volumeyear{year}
\volumenumber{number}
\issuenumber{number}
\eid{identifier}
\received[Received text]{date}

\revised[Revised text]{date}

\accepted[Accepted text]{date}

\published[Published text]{date}

\startpage{2}
\endpage{ }

The film growth governed by nucleation and growth is characterized by the
emergence of the scale invariance for some relevant quantities linked to the
islands forming the film, namely the size and the radial distribution
function. In this article we will place emphasis on the former.

Thanks to thorough numerical simulations of two dimensional (2D) thin film
growth for very large $R=D/F{a}^{4}$ ratio (where $D$ is the diffusion
coefficient of the adatoms, $F$ is the flux of impinging atoms and $a$ is the
lattice constant), it was induttively established the following relation for
the size ($s$) distribution of islands in the low coverage ($\Theta$) regime
\cite{[Bartelt1]}\cite{[Bales]}\cite{[Amar_Family]}%

\begin{equation}
N_{s}(\Theta)=\frac{\Theta}{\langle s\rangle^{2}}f\left(  \frac{s}{\langle
s\rangle}\right)  , \label{scaling}%
\end{equation}

where $\langle s \rangle$ is the average island size and $f(x)$ is the scaling
function. The first experimental evidence of scale invariance and in turn the
corroboration of eqn.\ref{scaling}, was given by Stroscio and Pierce who
studied the initial stages of the 2D growth of Fe on Fe(001)\cite{[StroPi]}.
Subsequently, other confirmations came from the study of the strained
heteroepitaxial growth of InAs on GaAs(001) in the submonolayer regime
\cite{[Bressler]}\cite{[Bell]}. However, InAs on GaAs(001) grows in a
Stranski-Krastanov mode, so there exists a critical coverage after which the
initial epitaxial (2D) growth is followed by the formation of 3D coherent
islands or quantum dots (QDs). It is a rather substantiated fact that also the
size (i.e. the volume or the number of atoms making up the dot) distribution
of QDs satisfies eqn.\ref{scaling} \cite{[Ebiko]}\cite{[Krzy]}\cite{[Arci]}.

At least as far as its derivation is concerned, eqn.\ref{scaling} is quite an
empirical relation, often referred to as scaling ansatz, and as we have seen,
both experimental data and numerical simulations seem to support it. It owes
to Mulheran and Blackman (MB) \cite{[Mulheran1]}\cite{[Mulheran2]} if
eqn.\ref{scaling} is, in fact, much more than a simple phenomenological
equation, in that they gave a convincing, neat explanation of its
geometric/physical origin. Their argument is rather simple and centers on the
concept of capture zone.

To begin with let us consider $N_{0}$ points per unit area distributed
throughout a plane. To these it is possible to associate a Voronoi network
\cite{[libro]} that has the property to tessellate the plane; to each point is
associated a cell or polygon. The cells do not overlap one another and the
space belonging to the i-th cell is that closer to the i-th point than to any
other of the remaining points. This geometric description finds an obvious
physical counterpart in film growth: surface, nucleation centers and capture
zones in place of plane, points and Voronoi cells, respectively. This analogy,
first proposed by Venables and Bell, \cite{[Venables]} is based on the
conjecture that the adatoms which land in the i-th cell are captured, on
average, by the i-th nucleation center, because, by definition, it is the
closest one.

It is then easy to convince oneself that, if all the nucleation centers
becomes active simultaneously, the size (volume) distribution of islands will
resemble the size distribution of the areas of the Voronoi cells because, on
average, a particular island is made up by the atoms previously contained in
the corresponding Voronoi cell. Incidentally, this conclusion is apparently
independent of the dimensionality of the islands. It follows that, since the
number of islands (nucleation centers) does not vary with $\Theta$, provided
the change of the shape of the Voronoi cells due to the growth of islands
\cite{[Bartelt2]} is negligible, the size distribution of islands does not
change its shape for all the coverage for which islands can be well
approximated by dimensionless points. We got to the point: \emph{the scale
invariance is nothing but the invariance of the Voronoi network}. As such,
even in the simplest case of simultaneous heterogeneous nucleation, discussed
above, the scale invariance, in principle, is not an exact law because it
rests on the structureless island approximation. In the event, however, this
approximation is rather good at small coverage which is, after all, the regime
where the scaling behavior is verified.\cite{[Bartelt1]}\cite{[Bales]}%
\cite{[Amar_Family]}

What about non-simultaneous nucleation?

Let us take into account, for the sake of simplicity, the homogeneous
nucleation case where dimer is the stable island. Taking it that the growth is
negligible during the initial nucleation stage, the rate equations of the
process reads%

\begin{equation}
\left\{
\begin{array}
[c]{cl}
& \dot{n_{1}}\cong1-Rn_{1}^{2}\smallskip\\
& \dot{n}=2Rn_{1}^{2},
\end{array}
\right.  \label{RE}%
\end{equation}

where $n_{1}$ and $n$ are the number of monomers and stable islands per site,
respectively. Since usually $Rn_{1}^{2}\ll1$, \cite{[Bales]} it can be
neglected from the first equation in the system eqn.\ref{RE}, thus one ends up with%

\begin{equation}
\label{nuclea}\dot{n}\approx2R\Theta^{2}.
\end{equation}

Eqn.\ref{nuclea} points out that a large value of the parameter $R$ entails a
fast nucleation. In other words, a large $R$ would involve so a fast
nucleation that it could be completed, for all practical purpose, before the
growth could become considerable. Such a scenario strongly resemble a
simultaneous nucleation event and consequently is a warning sign of scale
invariance. As a matter of fact, this is another condition for the island size
distribution function to display scale invariance in numerical
simulations.\cite{[Bartelt1]}\cite{[Bales]}\cite{[Amar_Family]}

In conclusion, to the extent that the islands can be considered structureless
and as long as the nucleation rate is adequately fast, the associated Voronoi
network is quasi-invariant and, as a consequence, quasi-scale invariance follows.

In this letter we display experimental Atomic Force Microscopy (VEECO-Digital)
data which support the aforementioned conclusion. In particular, we studied
the scale invariance properties of InAs QDs on GaAs(001).

The details of the experimental set up are reported elsewhere.\cite{[Arci]}%
\cite{[Placi]} Here it is enough to remark that the entire kinetics was
determined by a single shot of InAs on the same GaAs substrate. Our procedure
allows us to obtain a rather fine sampling of the range of coverage
($\Delta\Theta=0.01$ ML) and to avoid the use of a different substrate for
each value of the coverage. The substrate was held at $500$ $%
\operatorname{{}^{\circ}{\rm C}}%
$ and the InAs flux was $F=0.029$ ML/s.

In what it follows we focus only on the so called large QDs. Small QDs are
immaterial for what we are going to draw here, because their number soon
becomes negligible. \cite{[Arci]}\cite{[Placi]} The nucleation and its rate
are shown in fig.1. Their behavior points out that the condition for a
possible scale invariance is fulfilled, because the nucleation terminates, in
fact, within a time interval much shorter than that necessary to cover the
entire surface. In point of fact, the scaled distributions of the QD volumes
collapse in a single curve as shown in fig.2a. What is more, the satisfactory
agreement between the distribution of the QD volumes and the corresponding
distribution of the Voronoi-cell areas displayed in figs.2b-e corroborates
MB's theory.

There is still one last subject to be broached: the functional form of the
scaling function. Until today there exist only empirical proposals
\cite{[Bartelt1]} \cite{[Amar_Family]} \cite{[Mulheran1]} \cite{[Mulheran2]},
so it is a question of deciding what is the most satisfactory among them, and
not only on the basis of a mere numerical utilitarian employ but also, we
would say, \emph{above all}, on the epistemological point of view. There can
be no doubt that MB's proposal is the preferable one for the simple reason
that it is direct consequence of the explanation of the distribution scale
invariance. \emph{The distribution function must be the same as that of the
Voronoi cells}. This problem has been studying for many years \cite{[libro]}
and basically it remains on the conjecture of Kiang \cite{[Kiang]} that, after
having solved the problem in 1D, extended his result (but this is a
conjecture) to the 2D case. Following Kiang, MB proposed the following scale function%

\begin{equation}
\label{gamma}f_{\beta}(x)=\frac{\beta^{\beta}}{\Gamma(\beta)}x^{\beta
-1}e^{-\beta x},
\end{equation}

where $\beta\in\mathbb{R}$ is a parameter and $\Gamma(x)$ is the Euler's gamma function.

In ref.\cite{[Mulheran2]} this choice is discussed in comparison to the
function proposed in ref.\cite{[Amar_Family]} where, in turn, the comparison
with the proposal of ref.\cite{[Bartelt1]} is performed. We have fitted the
experimental volume distributions both to eqn.\ref{gamma} and to the
distribution of ref.\cite{[Amar_Family]}, here reported for the sake of clearness%

\begin{equation}
\label{AFDF}f_{i}(x)=C_{i}x^{i}e^{-ia_{i}x^{1/a_{i}}},
\end{equation}

where $i$ is the number of atoms making up the critical nucleus, $C_{i}$ and
$a_{i}$ are constants). Although both functions return more than acceptable
fits, the $\chi^{2}$ test is in favor of eqn.\ref{gamma} in four cases out of
six. In fig.3 we have plotted the values of $\beta$ returned by fits as a
function of $\Theta$. As it appears $\beta$ is constant within errors in
observance of the scaling law. The average $\bar{\beta}\cong4.5$ indicates
that there exists a certain degree of spatial correlation among the QDs,
because poissonian Voronoi tessellation returns $\beta\sim3.5$%
.\cite{[Mulheran1]} In order to determine the correlation length we performed
a series of computer simulations. The substrate is represented by a
($3000\times3000$) square lattice, where we select 720 points ($N_{0}%
=8\times10^{-5}$ very close to the experimental value) according to the rule
that two points cannot lay closer than a distance $\xi$ (hardcore correlation
approximation \cite{[FanTom]}). Afterwards the Voronoi tessellation is
generated and its distribution function is fitted to eqn.\ref{gamma}. Similar
as the mean experimental $\beta$ is obtained for $\xi\cong30-35$ which
corresponds to $30-35$ $%
\operatorname{nm}%
$ in the real surface. In fig.4 two typical simulations together with their
fits are shown.

Finally, a remark is in order. In our previous article \cite{[Arci]} we
analyzed the scaling properties on the basis of eqn.\ref{AFDF}, obtaining, on
the whole, very good results. As we have discussed above, this is not
surprising because often the scattering of experimental data makes it hard to
distinguish the slight differences between eqn.\ref{gamma} and eqn.\ref{AFDF}.
The point is that according to eqn.\ref{AFDF}, we proposed a possible scenario
which, in the light of the MB scaling explanation, cannot be entirely correct.
In actual fact, we were led astray by the importance assumed by the parameter
$i$ in eqn.\ref{AFDF}, according to which a bell shape distribution function
cannot be compatible with $i=0$. As a consequence, we were led to postulate
the existence of a possible "\emph{artifact of the erosion process}%
"\cite{[Arci]} which is, in fact, an unnecessary requirement in the contest of
MB's theory. What can be maintained with a high degree of confidence is that
the nucleation takes place at the step edges, the step erosion brings about an
explosive nucleation and the QD formation proceeds by diffusion and
aggregation only[see fig.2]. It remains to expound the exponential
distribution disclosed by the small QDs \cite{[Arci]}, on that we are working
on.\pagebreak

\textbf{Figure Captions}

\textbf{Fig. 1} - Experimental nucleation function of InAs/GaAs(001) quantum
dots (dotted curve) and the nucleation rate (solid curve).

\textbf{Fig. 2} - Scaled distributions of the experimental island volume for
QDs in the range $1.60-2.04$ ML of InAs coverages (a). Comparison between the
distribution of the QD volumes and the corresponding distribution of the
Voronoi-cell areas for the following InAs coverages: 1.65 ML (b), 1.68 ML (c),
1.70 ML (d), 1.79 ML\ (e).

\textbf{Fig. 3} - Behavior of the parameter $\beta$ as a function of coverage
as obtained by fitting the experimental distributions of Voronoi polygon areas.

\textbf{Fig. 4 - }Typical distributions of simulated Voronoi polygon areas.
The simulations have been carried out on a $3000\times3000$ square lattice
with a density $N_{0}=8\times10^{-5}$ point-like correlated dots. Correlation
lengths $\xi=30$ (a) and $\xi=35$ (b) have been used. Data have been fitted to
eqn.\ref{gamma} (see the text).

\end{document}